\documentclass{JHEP3}
\usepackage{amsmath}
\usepackage{amssymb}
\usepackage{graphicx}
\usepackage{cite}
\usepackage{tabularx}
\title{Color Dynamics in External Fields}

\author{Paolo Cea \\ Dipartimento
Interateneo di Fisica, Universit\`a di Bari  and INFN - Sezione di Bari, \\
I-70126 Bari, Italy \\
E-mail:  \email{Paolo.Cea@ba.infn.it} }

\author{Leonardo Cosmai \\ INFN - Sezione di Bari, I-70126 Bari,
Italy\\
E-mail:  \email{Leonardo.Cosmai@ba.infn.it}}

\received{\today}

\abstract{
We investigate the vacuum dynamics of U(1), SU(2), and SU(3) lattice gauge theories
in presence of external (chromo)magnetic
fields, both in (3+1) and (2+1) dimensions.
We find that the critical coupling for the phase transition in compact U(1) gauge theory
is independent of the strength of an external magnetic field.
On the other hand we find that, both in (3+1) and (2+1) dimensions, the deconfinement temperature for SU(2) and SU(3)
gauge systems in a constant abelian chromomagnetic field decreases when the
strength of the applied field increases.
We conclude that the dependence of the deconfinement temperature on the strength
of an external constant chromomagnetic field is a peculiar feature of
non abelian gauge theories and could be useful to get insight into color confinement.
}


\preprint{BARI-TH 2005/505}

\begin{document}

\newcommand{\be}{\begin{equation}}
\newcommand{\ee}{\end{equation}}

\section{Introduction}
\label{Introduction}
Color confinement is still a puzzling problem not withstanding the  large mess
of numerical investigations aimed to understand the nature of the QCD vacuum.
Indeed, the mechanism that leads to color confinement remains an open question
despite intense lattice studies for nearly three decades.

According to a model conjectured long time  ago by
G.~'t~Hooft~\cite{tHooft:1976eps} and S.~Mandelstam~\cite{Mandelstam:1974pi}
the confining vacuum behaves as a coherent state of color magnetic monopoles,
or, equivalently, the vacuum resembles a magnetic (dual) superconductor. Up to
now there is  numerical
evidence~\cite{Wosiek:1987kx,DiGiacomo:1990hc,Cea:1993sd,Singh:1993ma,Matsubara:1994nq,Bali:1995de,Cea:1995ed,Cea:1995zt,DiGiacomo:1996ca,Suzuki:2004dw}
in favor of chromoelectric flux tubes in pure lattice gauge vacuum. As well
there have been extensive numerical
studies~\cite{Stack:1992zp,Shiba:1995db,Arasaki:1997sm,Nakamura:1997sw,Chernodub:1997ps,Jersak:1999nv,DiGiacomo:1999fa,DiGiacomo:1999fb,Cea:1999zv,Hoelbling:2000su,Cea:2000zr,Carmona:2001ja}
of monopole condensation.

An alternative model for color confinement
is based on the special role of center vortices and $Z_N$ symmetry
(see~\cite{Greensite:2003bk} and references therein), even if the connection
of center symmetry to confinement has been recently
questioned in~\cite{Holland:2003kg,Holland:2003jy}.

One may conclude that there is no totally convincing explanation
of the confinement phenomenon
(for recent reviews on confinement see~\cite{Ripka:2003vv,Greensite:2003bk,Haymaker:1998cw})
and that a full understanding of the QCD vacuum dynamics  is still lacking. Indeed, as
recently observed~\cite{'tHooft:2004th} in connection with dual superconductivity picture,
even if magnetic monopoles do condense in
the confinement mode, the
actual mechanism of confinement could  depend on additional dynamical
forces. Therefore we feel that it is important to explore
any new paths that possibly may suggest new hints for understanding the
QCD vacuum.

In a previous paper~\cite{Cea:2001an}  we reported numerical results
showing that in four
dimensional U(1) lattice gauge theory the confining vacuum behaves
as a coherent condensate of Dirac magnetic monopoles, according
to analytical results in the literature~\cite{Fradkin:1978th}.
In the same paper we gave account of numerical results indicating
condensation of abelian magnetic monopoles and abelian vortices in
the confined phase of finite temperature SU(2) and SU(3) lattice gauge theories
in (3+1) dimensions.
Therefore one might conclude that in SU(2) and SU(3) gauge theories the
confining vacuum behaves as a coherent abelian magnetic condensate.
We also found~\cite{Cea:1999gn} that a weak constant
abelian chromomagnetic field at zero temperature is completely
screened in the continuum limit, while  at finite
temperature~\cite{Cea:2001pc,Cea:2002wx} our numerical results
indicate  that the applied field is restored by increasing the
temperature. These results strongly suggested that the confinement
dynamics could be intimately related to abelian chromomagnetic gauge
configurations.
Similar arguments have been reported in ref.~\cite{Skalozub:2002da}.
Moreover, in Refs.~\cite{Cea:2001pc,Cea:2002wx} the
SU(3) vacuum was probed by means of an external constant abelian
chromomagnetic field with increasing field strength. Remarkably, we
found that by increasing the strength of the applied external field
the deconfinement temperature decreases towards zero.  This means
that strong enough abelian chromomagnetic fields destroy the confinement
of color. In analogy to what happens in familiar superconductors when the strength of
an external magnetic field is increased (see for
instance ref.~\cite{Tinkham:1975}), this effect can be named
"reversible color Meissner effect" .
Altough the  existence of a critical chromomagnetic field is not
easily understandable  within the coherent magnetic monopole
condensate picture of the confining vacuum, it
could be directly explained if the vacuum behaves as an ordinary
relativistic color superconductor, or differently stated, if
the confining
vacuum resembles as a coherent condensate of tachionic color charged
scalar fields. Thus we
have to reconcile two apparently different aspects. From one hand,
the confining vacuum does display condensation of both  abelian
magnetic monopoles and vortices, on the other hand the relation
between the deconfinement temperature and the applied abelian
chromomagnetic field suggested that the vacuum behaves as a
condensate of an effective color charged scalar field whose mass is
proportional to the inverse of the magnetic
length~\cite{Cea:2001pc,Cea:2002wx}. The reversible color
Meissner effect  could be in agreement with R.~P.~Feynman who, in a
seminal paper~\cite{Feynman:1981ss},  argued that in three
dimensional SU(2) gauge theory long range correlation between
gluonic degrees of freedom destroys confinement. We would like to point out, to avoid misunderstanding,
that our reversible color Meissner effect is not related to color superconductivity
in cold dense quark matter (for a recent
review see ref.~\cite{Casalbuoni:2003wh} and references therein).

The aim of the present paper  is to investigate if our reversible
color Meissner effect is a generic feature
of non abelian gauge theories. To this end, we shall compare SU(3)
and SU(2) gauge theories in an external abelian chromomagnetic field
both in (3+1) and (2+1) dimensions. We shall, also, consider three
and four dimensional U(1) gauge theories in a magnetic background
field.

The plan of the paper is as follows. In sect.~\ref{gaugeinvariant} we briefly
recall for reader convenience our proposal of lattice effective action and
define the abelian chromomagnetic field on the lattice.
In sect.~\ref{4dim} we
present our results on vacuum dynamics in an external chromomagnetic background field
in (3+1)-dimensions for SU(3) and SU(2) at finite temperature,
and for U(1) at zero temperature.
Sect.~\ref{3dim} is devoted to corresponding results in (2+1)-dimensions.
Finally in sect.~\ref{Conclusions} we summarize and conclude. In
Appendix~\ref{su3T8}  we present results for SU(3) in an external
chromomagnetic background field directed along the direction $\hat{8}$ in color
space.
%
\section{The gauge invariant  lattice effective action }
\label{gaugeinvariant}
In our previous studies, in order to investigate vacuum structure of  lattice
gauge theories both at zero and finite temperature, we introduced a lattice
effective action for gauge systems in external static background fields. In
this section, for reader convenience, we shall briefly summarize our proposal
of lattice effective action which is gauge invariant against static gauge
transformations of the background field.
\subsection{The lattice effective action: $T=0$}
\label{effaction}
In Refs.~\cite{Cea:1997ff,Cea:1999gn} we introduced a lattice gauge invariant
effective action $\Gamma[\vec{A}^{\text{ext}}]$ for an external background
field $\vec{A}^{\text{ext}}$:
\be
\label{Gamma}
\Gamma[\vec{A}^{\text{ext}}] = -\frac{1}{L_t} \ln
\left\{
\frac{{\mathcal{Z}}[\vec{A}^{\text{ext}}]}{{\mathcal{Z}}[0]}
\right\}
\ee
where $L_t$ is the lattice size in time direction and
$\vec{A}^{\text{ext}}(\vec{x})$ is the continuum gauge potential of the
external static background field.  ${\mathcal{Z}}[\vec{A}^{\text{ext}}]$ is the
lattice partition functional
\be \label{Zetalatt}
{\mathcal{Z}}[\vec{A}^{\text{ext}}] =
\int_{U_k(\vec{x},x_t=0)=U_k^{\text{ext}}(\vec{x})} {\mathcal{D}}U \; e^{-S_W}
\,, \ee
with $S_W$ the standard pure gauge Wilson action.

The functional integration is performed over the lattice links, but constraining
the spatial links belonging to a given time slice (say $x_t=0$) to be
\be
\label{coldwall}
U_k(\vec{x},x_t=0) = U^{\text{ext}}_k(\vec{x})
\,,\,\,\,\,\, (k=1,2,3) \,\,,
\ee
 $U^{\text{ext}}_k(\vec{x})$ being the lattice version  of the external continuum
gauge potential $\vec{A}^{\text{ext}}(x)=\vec{A}^{\text{ext}}_a(x) \lambda_a/2$.
Note that the temporal links are not constrained.

In the case of a static background field which does not vanish at infinity we
must also impose that, for each time slice $x_t \ne 0$, spatial links exiting
from sites belonging to the spatial boundaries  are fixed according to
eq.~(\ref{coldwall}). In the continuum this last condition amounts to the
requirement that fluctuations over the background field vanish at infinity.

The partition function defined in eq.~(\ref{Zetalatt}) is also known as lattice
Schr\"odinger functional~\cite{Luscher:1992an,Luscher:1995vs} and in the
continuum corresponds to the Feynman kernel~\cite{Rossi:1980jf}. Note that, at
variance with the usual formulation of the lattice Schr\"odinger
functional~\cite{Luscher:1992an,Luscher:1995vs} where a lattice cylindrical
geometry is adopted, our lattice has an hypertoroidal geometry so that $S_W$ in
eq.~(\ref{Zetalatt}) is allowed to be the standard Wilson action.

The lattice effective action $\Gamma[\vec{A}^{\text{ext}}]$ corresponds to the vacuum
energy, $E_0[\vec{A}^{\text{ext}}]$,
in presence of the background field with respect to the vacuum energy, $E_0[0]$,
with  $\vec{A}^{\text{ext}}=0$
\be
\label{vacuumenergy}
\Gamma[\vec{A}^{\text{ext}}] \quad \longrightarrow \quad E_0[\vec{A}^{\text{ext}}]-E_0[0] \,.
\ee
The relation above is true by letting the temporal lattice size $L_t \to \infty$;
on finite lattices this amounts to have $L_t$ sufficiently large
to single out the ground state contribution to the energy.

Since the lattice effective action eq.~(\ref{Gamma}) is given in terms of the
lattice Schr\"odinger functional, which is invariant for time-independent gauge
transformation of the background field~\cite{Luscher:1992an,Luscher:1995vs}, it
is gauge invariant too.

\subsection{The thermal partition functional}
\label{thermalpf}

If we now consider the gauge theory at finite temperature $T=1/(a L_t)$
in presence of an external background field, the relevant quantity turns out to be
the free energy functional defined as
\be
\label{freeenergy}
{\mathcal{F}}[\vec{A}^{\text{ext}}] = -\frac{1}{L_t} \ln
\left\{
\frac{{\mathcal{Z_T}}[\vec{A}^{\text{ext}}]}{{\mathcal{Z_T}}[0]}
\right\} \; .
\ee
${\mathcal{Z_T}}[\vec{A}^{\text{ext}}]$ is the thermal partition
functional~\cite{Gross:1981br}
in presence of the background field $\vec{A}^{\text{ext}}$, and is defined as
\be
\label{ZetaTnew}
\mathcal{Z}_T \left[ \vec{A}^{\text{ext}} \right]
= \int_{U_k(\vec{x},L_t)=U_k(\vec{x},0)=U^{\text{ext}}_k(\vec{x})}
\mathcal{D}U \, e^{-S_W}   \,.
\ee
In eq.~(\ref{ZetaTnew}), as in eq.~(\ref{Zetalatt}), the spatial links
belonging to the time slice $x_t=0$ are constrained to the value of the
external background field, the temporal links are not constrained.  On a
lattice with finite spatial extension we also  usually impose that the links at
the spatial boundaries are fixed according to boundary conditions
eq.~(\ref{coldwall}),  apart from the case in which the external background
field vanishes at spatial infinity (as happens for the monopole field), where
the choice of periodic boundary conditions in the spatial direction is
equivalent to  eq.~(\ref{coldwall}) in the thermodynamical limit. If the
physical temperature is sent to zero, the thermal functional
eq.~(\ref{ZetaTnew}) reduces to the zero-temperature Schr\"odinger functional
eq.~(\ref{Zetalatt}). The free energy functional eq.~(\ref{freeenergy})
corresponds to the free energy, $F[\vec{A}^{\text{ext}}]$, in presence of the
external background field evaluated with respect to the free energy, $F[0]$,
with $\vec{A}^{\text{ext}}=0$. When the physical temperature is sent to zero
the free energy  functional reduces to the vacuum energy functional
eq.~(\ref{Gamma}).

\subsection{Abelian (chromo)magnetic background field}
\label{abelianfield}

We are interested in vacuum dynamics of  U(1), SU(2), and SU(3) lattice
gauge theories under the influence of an
abelian chromomagnetic background field.

In our previous studies we found that in SU(2) and SU(3)
at zero temperature a (not too strong) constant abelian chromomagnetic field
at zero temperature is completely screened
in the continuum limit~\cite{Cea:1999gn}.
We also found that in SU(3) the deconfinement temperature depends on the strength of an applied external
constant abelian chromomagnetic field~\cite{Cea:2002wx}.
This is at variance of abelian magnetic monopoles where the abelian monopole background fields
do not modify the deconfinement temperature~\cite{Cea:2004ux}.
We would like to corroborate our findings with further investigations,
in particular we would like to ascertain if the dependence of the deconfinement
temperature on the strength of an applied external
constant abelian chromomagnetic field is a peculiar feature of non abelian gauge theories.

Let us now define a static constant abelian chromomagnetic field on the lattice.
We first consider the SU(3) case.
In the continuum the gauge potential giving rise to a static constant abelian chromomagnetic field
directed along spatial direction $\hat{3}$ and direction $\tilde{a}$ in the color space
is given by
\be
\label{su3pot}
\vec{A}^{\text{ext}}_a(\vec{x}) =
\vec{A}^{\text{ext}}(\vec{x}) \delta_{a,\tilde{a}} \,, \quad
A^{\text{ext}}_k(\vec{x}) =  \delta_{k,2} x_1 H \,.
\ee
In SU(3) lattice gauge theory
the constrained lattice links (see eq.~(\ref{coldwall})) corresponding to
the continuum gauge potential eq.~(\ref{su3pot}) are (choosing $\tilde{a}=3$, i.e. abelian
chromomagnetic field along direction $\hat{3}$ in color space)
\be
\label{t3links}
\begin{split}
& U^{\text{ext}}_1(\vec{x}) =
U^{\text{ext}}_3(\vec{x}) = {\mathbf{1}} \,,
\\
& U^{\text{ext}}_2(\vec{x}) =
\begin{bmatrix}
\exp(i \frac {g H x_1} {2})  & 0 & 0 \\ 0 &  \exp(- i \frac {g H
x_1} {2}) & 0
\\ 0 & 0 & 1
\end{bmatrix}
\,.
\end{split}
\ee
We will refer to this case as $T_3$ abelian chromomagnetic field.
If we choose instead abelian
chromomagnetic field along direction $\hat{8}$ in color space the constrained
lattice links are given by
\be
\label{t8links}
\begin{split}
& U^{\text{ext}}_1(\vec{x}) =
U^{\text{ext}}_3(\vec{x}) = {\mathbf{1}}
\\
& U^{\text{ext}}_2(\vec{x}) =
\begin{bmatrix}
\exp(i \frac {g H x_1} {2 \sqrt{3}})  & 0 & 0 \\ 0 &  \exp(i \frac {g H
x_1} {2 \sqrt{3}}) & 0
\\ 0 & 0 & \exp(-i \frac {g H x_1} {\sqrt{3}})
\end{bmatrix}
\,.
\end{split}
\ee
We will refer to this case as $T_8$ abelian chromomagnetic field.
Since our lattice has the topology of a torus,
the magnetic field turns out to be quantized
\be
\label{quant} a^2 \frac{g H}{2} = \frac{2 \pi}{L_1}
n_{\text{ext}} \,, \qquad  n_{\text{ext}}\,\,\,{\text{integer}}\,.
\ee
In the case of SU(2) lattice gauge theories the constrained spatial
links are
\be
\label{su2links}
\begin{split}
& U^{\text{ext}}_1(\vec{x}) = U^{\text{ext}}_3(\vec{x}) = {\mathbf{1}} \,, \\
& U^{\text{ext}}_2(\vec{x}) = \cos( \frac{ gHx_1}{2} ) +
i \sigma^3 \sin( \frac{gHx_1}{2} )   \,,
\end{split}
\ee
$\sigma^3$ being the Pauli matrix.\\
Finally in the U(1) case the constrained spatial links corresponding
to a constant magnetic background field (along spatial direction $\hat{3}$) are
\be
\label{u1links}
\begin{split}
& U^{\text{ext}}_1(\vec{x}) = U^{\text{ext}}_3(\vec{x}) = {\mathbf{1}} \,, \\
& U^{\text{ext}}_2(\vec{x}) = \cos(  gHx_1) +
i \sin( gHx_1 )   \,.
\end{split}
\ee
Since the free energy functional $F[\vec{A}^{\text{ext}}]$
is invariant for time independent gauge transformations of
the background field $\vec{A}^{\text{ext}}$, it follows that
for a constant background field, $F[\vec{A}^{\text{ext}}]$
is proportional to the spatial volume $V=L_s^3$, and the relevant
quantity is the density $f[\vec{A}^{\text{ext}}]$ of
free energy
\be
\label{free-energy} f[\vec{A}^{\text{ext}}] = \frac{1}{V}
F[\vec{A}^{\text{ext}}] \,.
\ee
We evaluate by numerical simulations the
derivative with respect to the coupling $\beta$
of the free energy density $f[\vec{A}^{\text{ext}}]$ at fixed
external field strength $gH$
\be
\label{deriv}
f^{\prime}[\vec{A}^{\text{ext}}]    = \left \langle
\frac{1}{\Omega} \sum_{x,\mu < \nu}
\frac{1}{3} \,  \text{Re}\, {\text{Tr}}\, U_{\mu\nu}(x) \right\rangle_0  \\
  - \left\langle \frac{1}{\Omega} \sum_{x,\mu< \nu} \frac{1}{3} \,  \text{Re} \, {\text{Tr}} \, U_{\mu\nu}(x)
\right\rangle_{\vec{A}^{\text{ext}}} \,,
\ee
where the subscripts on the averages indicate the value of the external field
and $\Omega = L_s^3 \times L_t$ is the lattice volume.
The generic plaquette
$U_{\mu\nu}(x)=U_\mu(x)U_\nu(x+\hat{\mu})U^\dagger_\mu(x+\hat{\nu})U^\dagger_\nu(x)$
contributes to the sum in eq.~(\ref{deriv}) if the link $U_\mu(x)$
is a "dynamical" one, i.e. it is not constrained in the functional
integration eq.~(\ref{ZetaTnew}).
Observing that $f[\vec{A}^{\text{ext}}] = 0$ at $ \beta = 0$, we
may eventually obtain  $f[\vec{A}^{\text{ext}}]$ from
$f^{\prime}[\vec{A}^{\text{ext}}]$ by numerical integration:
\be
\label{trapezu1}
f[\vec{A}^{\text{ext}}]  =  \int_0^\beta
f^{\prime}[\vec{A}^{\text{ext}}] \,d\beta^{\prime} \; .
\ee
%

\section{(3+1) dimensions}
\label{4dim}
In this section we report results obtained in studying the finite temperature
phase transition of lattice gauge theories SU(3) and  SU(2) in (3+1)-dimensions,
in presence of a constant abelian chromomagnetic background field.
We shall also report results for confinement-Coulomb phase transition
in U(1) lattice gauge theory at zero temperature in a constant magnetic background field.
A preliminary account of our results has ben presented in ref.~\cite{Cea:2004rk}.

\subsection{SU(3)}
\label{4dimSU3}
We simulate SU(3) pure gauge theory in a constant abelian background field
defined in Eqs.~(\ref{su3pot}) and~(\ref{t3links}).
As is well known, the pure SU(3) gauge system undergoes a deconfinement
phase transition at a given critical temperature.
Our aim is to study the possible dependence of the critical temperature from
the strength of the applied field.
The critical coupling $\beta_c$ can be evaluated by measuring $f^{\prime}[\vec{A}^{\text{ext}}]$,
the derivative of the free energy density with respect to $\beta$,
as a function of $\beta$.
Indeed we found that $f^{\prime}[\vec{A}^{\text{ext}}]$ (see eq.~(\ref{deriv})) displays
a peak in the critical region (see fig.~\ref{Fig1}) where it can be parameterized as
\be
\label{peak-form}
\frac{f^{\prime}(\beta,L_t)}{\varepsilon^{\prime}_{\text{ext}}}
= \frac{a_1(L_t)}{a_2(L_t) [\beta - \beta^*(L_t)]^2 +1} \,.
\ee
In eq.~(\ref{peak-form}) we normalize $f^{\prime}$  to
$ \varepsilon^{\prime}_{\text{ext}}$,
the  derivative of the classical energy due to  the external
applied field
\be
\label{classical}
\varepsilon^{\prime}_{\text{ext}} =
\frac{2}{3} \, [1 - \cos( \frac{g H}{2} )] = \frac{2}{3} \, [1 -
\cos( \frac{2 \pi}{L_1} n_{\text{ext}})]  \,.
\ee
%
%
\FIGURE[ht]{\label{Fig1}
\includegraphics[width=0.85\textwidth,clip]{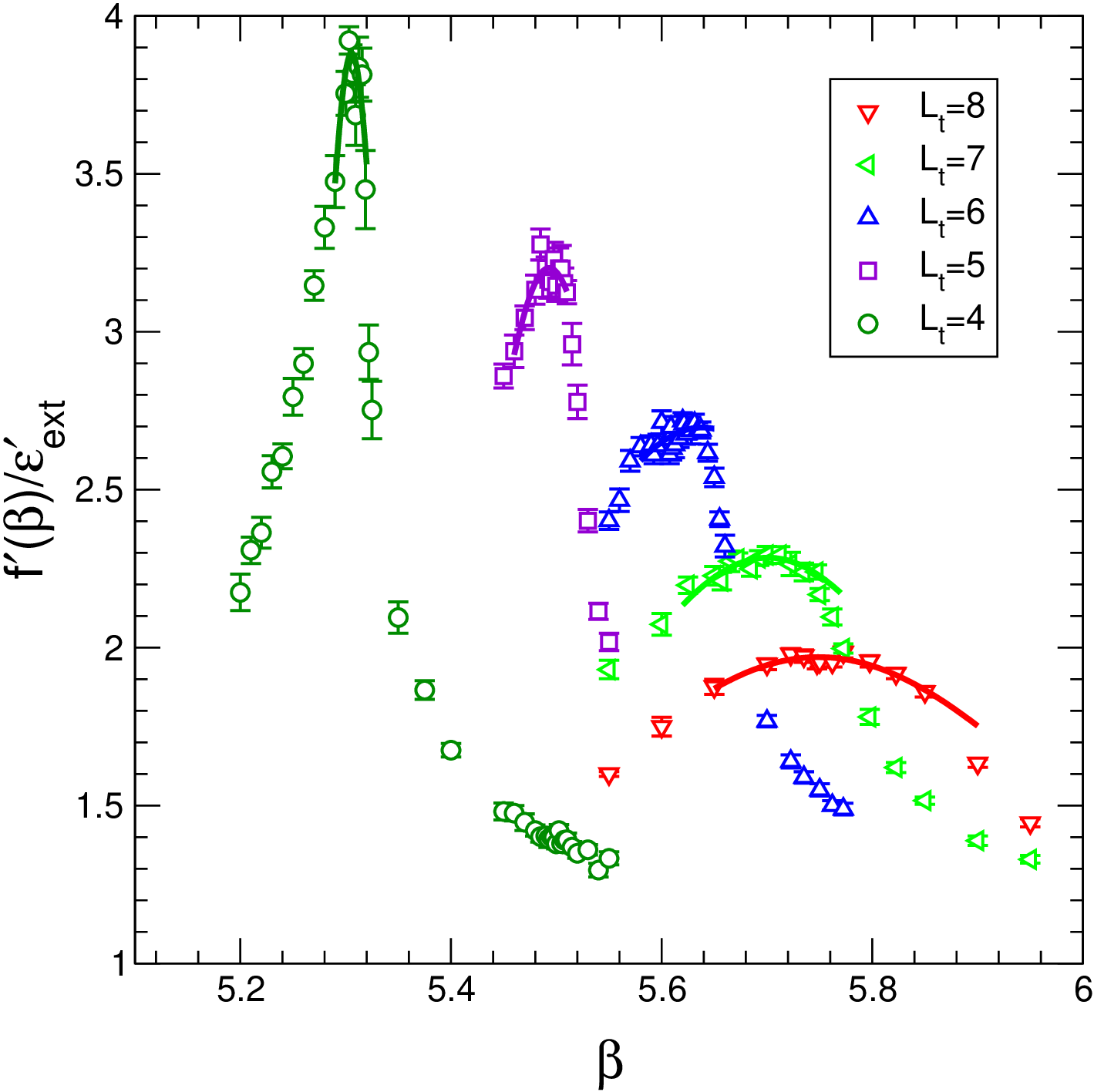}
\caption{SU(3) in (3+1) dimensions. The derivative of the free energy density with respect to
the gauge coupling $\beta$, eq.~(\ref{deriv}), versus $\beta$ at fixed external field
strength ($n_{\text{ext}}=1$) for spatial lattice size $L_s=64$ and
temporal lattice sizes $L_t=4,5,6,7,8$. Solid lines are the fits eq.~(\ref{peak-form}).}
}
Remarkably, we have checked that the evaluation of the critical coupling
$\beta^*(L_t)$ by means of $f^{\prime}[\vec{A}^{\text{ext}}]$ is
consistent with the usual determination obtained through the temporal
Polyakov loop susceptibility:
\be
\label{Polysuscep}
\begin{split}
&  \chi(|P|) = <|P|^2> - <|P|>^2 \\
& P = \frac{1}{V_s} \sum_{\vec{x}} \frac{1}{3} \, {\text{Tr}}  \prod_{x_4=1}^{L_t} U_4(x_4,\vec{x}) \,.
\end{split}
\ee
The Polyakov loop susceptibility near the peak has been obtained by
means of the density spectral method~\cite{Ferrenberg:1989ui,Newman:1999}.
The statistical errors for the points belonging to the extrapolated curve near the peak,
as well as the position of the peak and its statistical error, were evaluated by means
of a bootstrap analysis~\cite{Davison:1997}.
For instance, on
a $64^3 \times 6$ lattice and $ n_{\text{ext}}=1$ we get $
\beta_c=5.6272(69)$ from eq.~(\ref{peak-form}) and $
\beta_c=5.6266(12)$ when evaluating the peak of the Polyakov loop
susceptibility by means of the density spectral method  (see fig.~\ref{Fig2}).
%
%
%
\FIGURE[ht]{\label{Fig2}
\includegraphics[width=0.85\textwidth,clip]{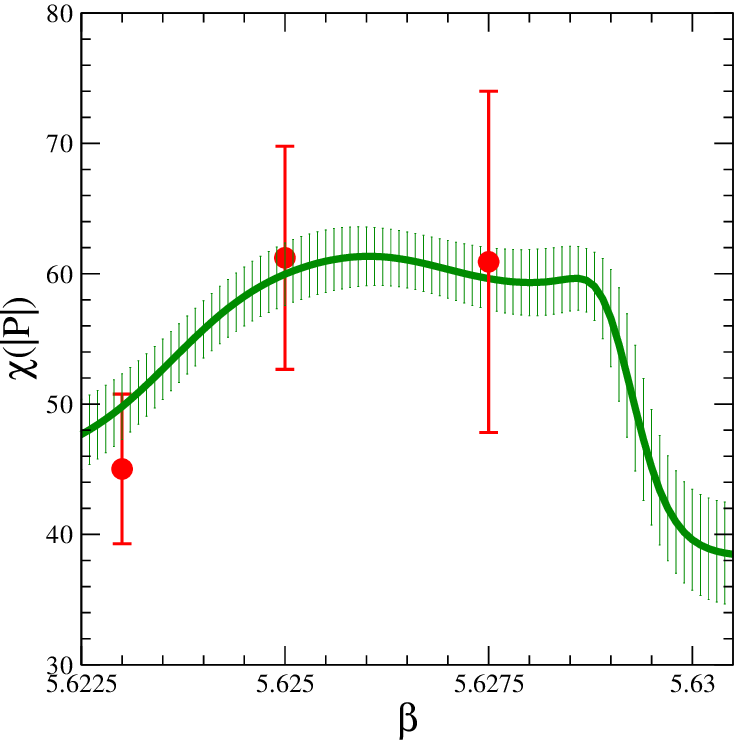}
\caption{SU(3) in (3+1) dimensions. The susceptibility of the absolute value of the Polyakov loop evaluated for three values
of $\beta$ on a $64^3 \times 6$ lattice and external field strength $n_{\text{ext}}=1$.
The solid line has been obtained by reweighting. The error bars have been estimated
using the bootstrap method.}
}

Once $\beta^*(L_t)$ has been determined, the deconfinement
temperature can be preliminarily estimated in units of $\Lambda_{\text{latt}}$.
Indeed
\be
\label{Tc}
\frac{T_c}{\Lambda_{\text{latt}}} = \frac{1}{L_t}
\frac{1}{f_{SU(3)}(\beta^*(L_t))} \,,
\ee
where
\be
\label{fsun}
f_{SU(N)}(\beta) = \left( \frac{\beta}{2  N b_0}
\right)^{b_1/2b_0^2} \, \exp \left( -\beta \frac{1}{4 N b_0}
\right) \,,
\ee
$N$ being the color number, $b_0=(11N)/(48 \pi^2)$, and $b_1=(34 N^2)/(3(16 \pi^2)^2)$.
In order to obtain the continuum limit of the critical temperature we
have to extrapolate
$ T_c/\Lambda_{\text{latt}}$, given by eq.~(\ref{Tc}),
to the continuum. This can be done, following ref.~\cite{Fingberg:1993ju}, by means of a linear extrapolation of $T_c/\Lambda_{\text{latt}}$
as a function of $a T_c$ for $a T_c \to 0$.
We varied the strength of the applied external
abelian chromomagnetic background field to study quantitatively
the dependence of $ T_c$ on $gH$.
%
%
%
%
%
\FIGURE[ht]{\label{Fig3}
\includegraphics[width=0.85\textwidth,clip]{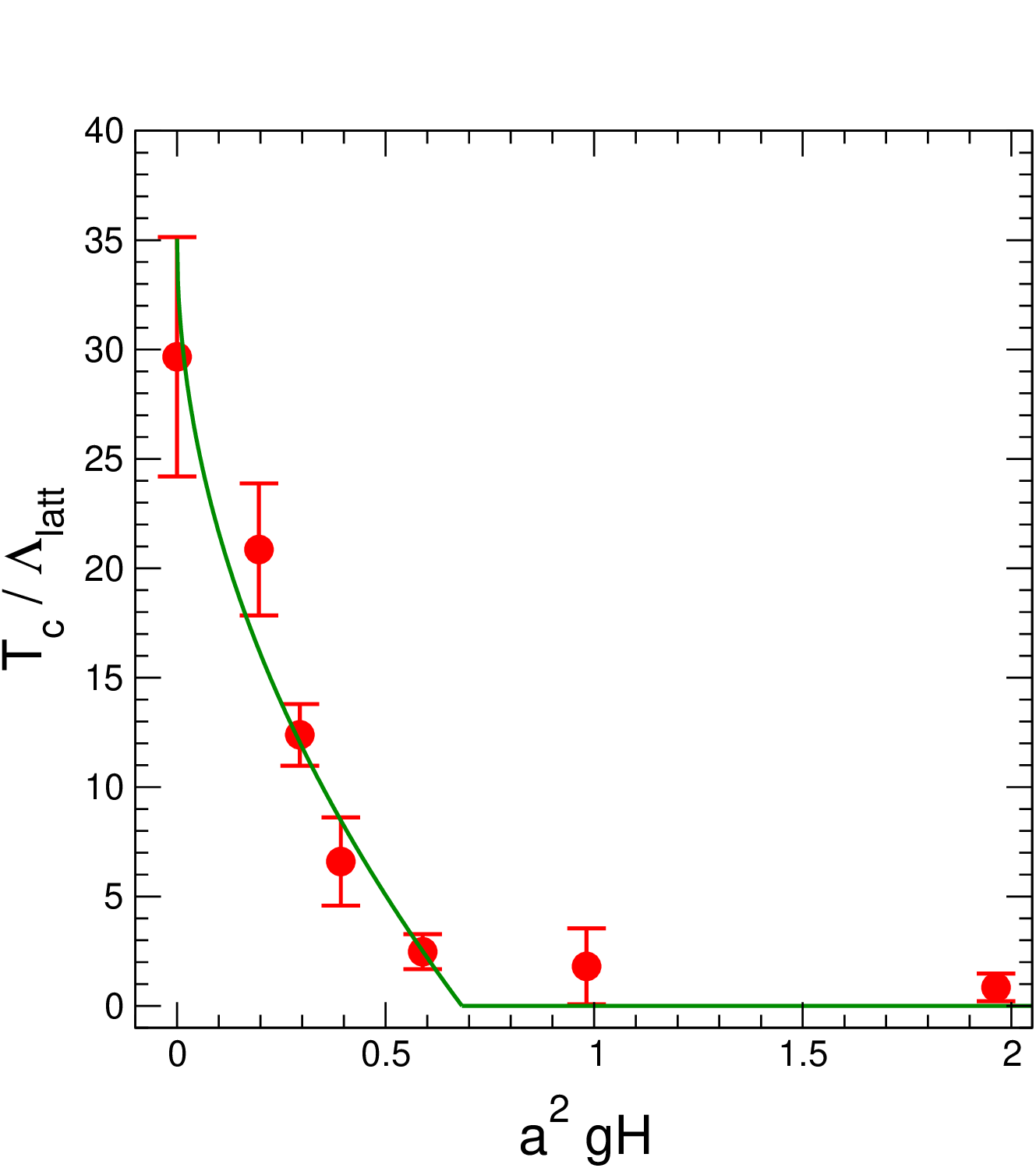}
\caption{SU(3) in (3+1) dimensions. The continuum critical temperature $T_c$ in units of $\Lambda_{\text{latt}}$ versus the
external field strength eq.~(\ref{quant}) (in lattice units). Solid line is the fit
eq.~(\ref{Tcfit}) to our data.}
}
%
%
%
%
\FIGURE[ht]{\label{Fig4}
\includegraphics[width=0.85\textwidth,clip]{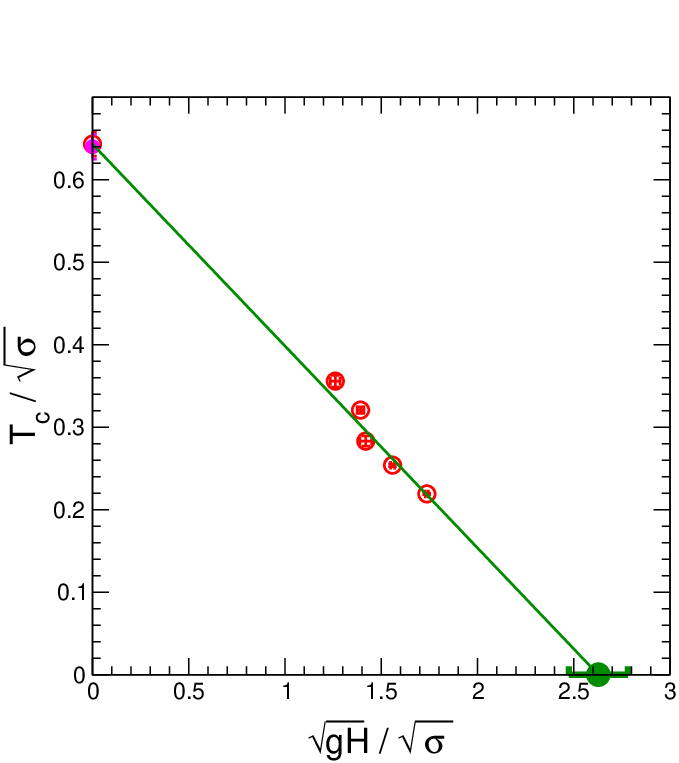}
\caption{SU(3) in (3+1) dimensions. The critical temperature $T_c$ estimated on a $64^3 \times 8$ lattice in units of
the string tension, eq.~(\ref{Tphys}), versus the square root of the field strength
$\sqrt{gH}$ in units of the string tension, eq.~(\ref{gHphys}). Solid line is the linear fit eq.~(\ref{tcsqrtsigma}).
In correspondence of zero vertical axis: open circle  is $T_c/\sqrt{\sigma}$ at zero external field, Eq.~(\ref{tczero});
full circle is the determination of $T_c/\sqrt{\sigma}$ obtained in the literature~\cite{Teper:1998kw}
The green full point is the critical field in units of the string tension, eq.~(\ref{su3hc}).
}
}
From fig.~\ref{Fig3}, where we display $T_c/\Lambda_{\text{latt}}$ versus $gH$ in
lattice units,  we may conclude that the critical
temperature decreases by increasing the strength of the external abelian
chromomagnetic field and eventually goes to zero for a strong enough external field.

To get more insight into this result we can try to parameterize the
behavior of the critical temperature versus the applied field strength.
As a matter of fact, if the magnetic length, defined as
$a_H \sim 1/\sqrt{gH}$, is the only relevant
scale of the problem for dimensional reasons one expects that
\be
\label{dimensional-form}
T_c^2 \, \, \sim \,  \, gH \, .
\ee
Indeed, as fig.~\ref{Fig3} displays, we get a good fit to our data
using the following parameterization
\be
\label{Tcfit}
\frac{T_c(gH)}{\Lambda_{\text{latt}}} =
\frac{T_c(0)}{\Lambda_{\text{latt}}} + \alpha \sqrt{a^2gH} \,.
\ee
We get
\be
\label{fittc}
\begin{split}
& T_c(0)/\Lambda_{\text{latt}}=35.5\pm5.2 \\
& \alpha=-42.4 \pm 7.4 \,.
\end{split}
\ee
It is worthwhile to note that our estimation for $T_c(0)/\Lambda_{\text{latt}}$
is compatible with $T_c(0)/\Lambda_{\text{latt}}=29.67\pm5.47$
obtained in ref.~\cite{Fingberg:1993ju} with completely different methods.

The preliminary analysis of our lattice data drives us to conclude
that, remarkably, a critical field,
$gH_c \simeq 0.68$ (in lattice units), exists
such that $ T_c=0$ for $ gH>gH_c$.
This kind of behavior could be interpreted as the colored counterpart of the Meissner effect
in ordinary superconductors, when strong enough magnetic fields
destroy the superconductive BCS vacuum~\cite{Tinkham:1975}.
Then we shall refer to this remarkable result as the reversible color Meissner effect.
Once again we would like to stress that this effect is not related to the color superconductivity
in cold dense quark matter. Indeed, we believe  that our
reversible color Meissner effect  is deeply rooted in the non-perturbative
color confining nature of the vacuum and could be a window open towards unraveling
the true nature of the confining vacuum.

So far we reported our results for the critical temperature $T_c$ in units of $\Lambda_{\text{latt}}$
and for the critical strength of the abelian chromomagnetic background field in
lattice units.
However it is well known that asymptotic scaling could be affected by
scaling violation effects due to finite size of the lattice. On the other hand, such as
effects are strongly reduced in the scaling of physical quantities. So that
it is  useful to analyze our data in physical units.
In a pure gauge theory this can be done in terms of the string tension $\sigma$
computed at zero temperature in correspondence of the value of the gauge
coupling $\beta = \beta_c$.
We do not need to directly compute the string tension, for we may use the following
parameterization of the SU(3) string tension given by Edwards et al. (see
eq.~(4.4) in ref.~\cite{Edwards:1998xf})
\be
\label{sigma}
\begin{split}
& (a\sqrt{\sigma})(g)  =  \, f_{SU(3)}(g^2) ~ ( \,
  1 + 0.2731      \, \hat{a}^2(g)
    - 0.01545     \, \hat{a}^4(g)
    + 0.01975     \, \hat{a}^6(g) \, )/0.01364 \, , \\
&  \hat{a}(g) = \frac{f_{SU(3)}(g^2)}{f_{SU(3)}(g^2(\beta=6))} \,; \qquad \beta=\frac{6}{g^2} \,,
\end{split}
\ee
for $5.6 \leq \beta \leq 6.5$; $f_{SU(3)}$ is defined in eq.~(\ref{fsun}).
The critical temperature in physical units is given by
\be
\label{Tphys}
\frac{T_c}{\sqrt{\sigma(\beta_c)}} = \frac{1}{L_t \sqrt{\sigma(\beta_c)}} \,.
\ee
Moreover, using eq.~(\ref{quant}), the field strength  is
\be
\label{gHphys}
\frac{\sqrt{gH}}{\sqrt{\sigma(\beta_c)}} = \sqrt{\frac{4 \pi n_{\text{ext}}}{L_x  \sigma(\beta_c)}} \,.
\ee
Our data for $T_c/\sqrt{\sigma}$ versus $\sqrt{gH}/\sqrt{\sigma}$ on a
$64^3 \times 8$ lattice are displayed in fig.~\ref{Fig4}.
It is worth to note that, consistently with our previous analysis, lattice data can
be reproduced by the linear fit
\be
\label{tcsqrtsigma}
\frac{T_c}{\sqrt{\sigma}} =  \alpha \frac{\sqrt{gH}}{\sqrt{\sigma}} + \frac{T_c(0)}{\sqrt{\sigma}} \,,
\ee
with
\be
\label{tczero}
\frac{T_c(0)}{\sqrt{\sigma}} = 0.643(15)  \qquad  \alpha = -0.245(9)  \,.
\ee
It is noticeable that our determination for $T_c(0)/\sqrt{\sigma}$ is consistent with the determinations
$T_c/\sqrt{\sigma}=0.640(15)$  obtained
in the literature  without external field~\cite{Teper:1998kw}.
Using eq.~(\ref{tcsqrtsigma}) the critical field can now be expressed in units of the string tension
\be
\label{su3hc}
\frac{\sqrt{gH_c}}{\sqrt{\sigma}} = 2.63  \pm 0.15  \,.
\ee
Assuming $\sqrt{\sigma}=420$~MeV, eq.~(\ref{su3hc}) gives for the critical field
\be
\label{criticalfield}
\sqrt{gH_c} = (1.104 \pm 0.063) {\text{GeV}}
\ee
corresponding to $gH_c=6.26(2) \times 10^{19}$~Gauss.
Recently, it has been suggested that strong magnetic fields of order $10^{19}$~Gauss
are naturally associated with the QCD scale~\cite{Kabat:2002er}. Moreover, it is believed
that large magnetic fields might be generated during cosmological phase transitions.
So that, we see that our findings could imply interesting effects during
the cosmological deconfinement transition, which are worthwhile to investigate.
%
%
%
\FIGURE[ht]{\label{Fig5}
\includegraphics[width=0.85\textwidth,clip]{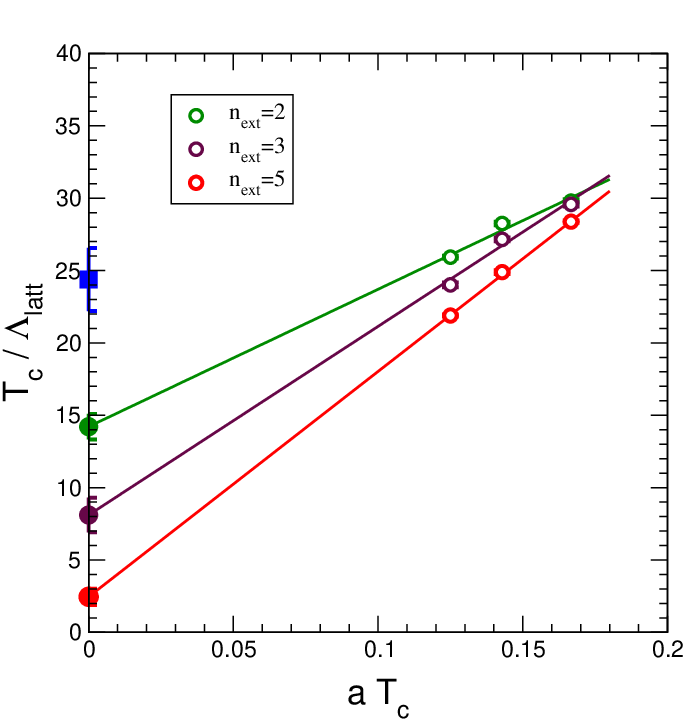}
\caption{SU(2) in (3+1) dimensions. $T_c/\Lambda_{\text{latt}}$ versus $aT_c$ for three different values of the
external field strength ($n_{\text{ext}}=2,3,5$) on  $64^3 \times L_t$ lattices ($L_t=6,7,8$). Full points in correspondence of $aT_c=0$ are
the extrapolations to the continuum. Full square is the critical temperature for SU(2) lattice
gauge theory without external field taken from ref.~\cite{Fingberg:1993ju}.}
}
%
%
%
%
%
%
%
\FIGURE[ht]{\label{Fig5a}
\includegraphics[width=0.85\textwidth,clip]{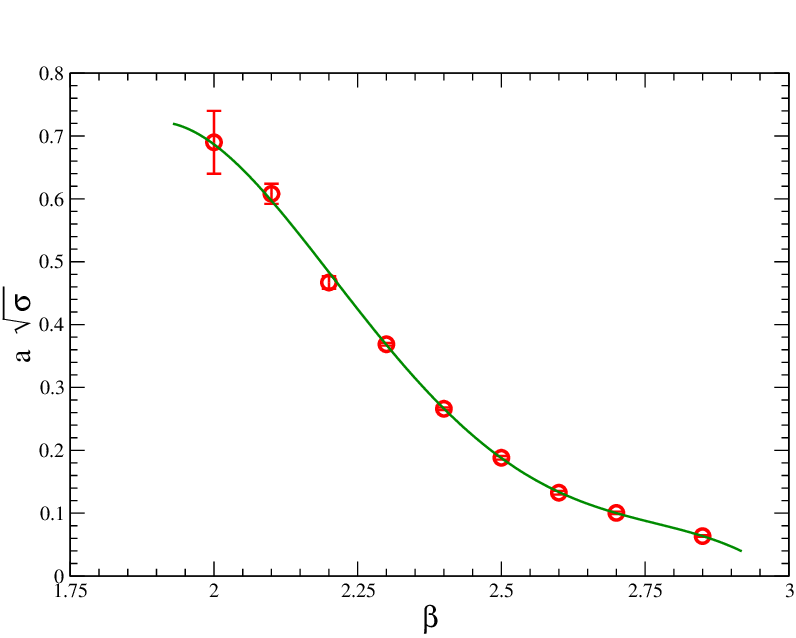}
\caption{SU(2) string tension in (3+1) dimensions. Open circles are taken from
Table~10 of ref.~\cite{Teper:1998kw}. The solid line is our best fit with
Chebyshev polynomials of the first kind up to order 6.}
}
%
%
%
%
%
%
%
\FIGURE[ht]{\label{Fig6}
\includegraphics[width=0.85\textwidth,clip]{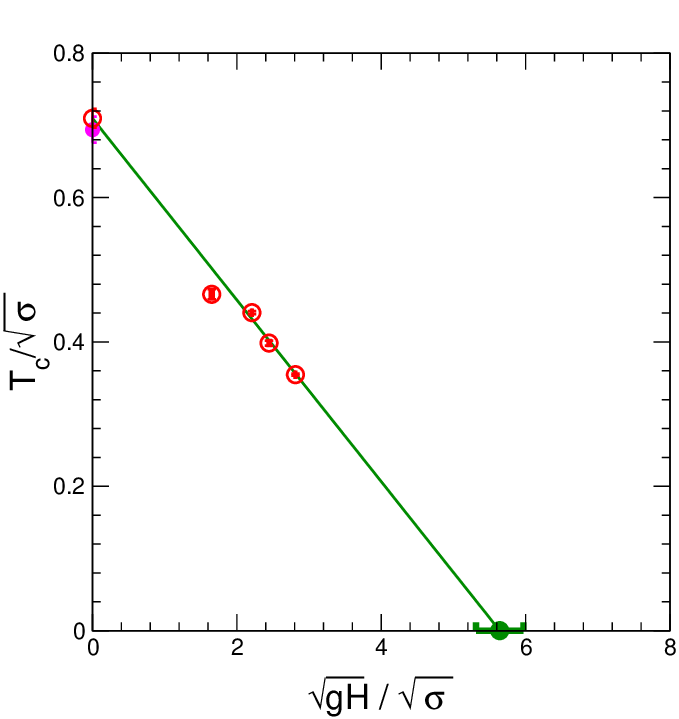}
\caption{SU(2) in (3+1) dimensions. The critical temperature $T_c$ estimated on a $64^3 \times 8$ lattice in units of
the string tension, eq.~(\ref{Tphys}), versus the square root of the field strength
$\sqrt{gH}$ in units of the string tension, eq.~(\ref{gHphys}). Solid line is the linear fit eq.~(\ref{tcsqrtsigma})
On the zero vertical axis are represented the extrapolation of our data to zero value of the field (open circle)
and the value for $T_c(0)/\sqrt{\sigma}$ (without external field) given in ref.~\cite{Teper:1998kw} (full circle).
The green full circle is the critical field in units of the string tension, eq.~(\ref{su2hc}).
}
}
%
%
%
%
%
\subsection{SU(2)}
\label{4dimSU2}
We also studied the SU(2) lattice gauge theory in a constant abelian chromomagnetic
field.
Even in this theory the
deconfinement temperature turns out to depend on the strength of the applied
chromomagnetic field, as already discussed in sect.~\ref{Introduction}.

We evaluated the critical coupling $\beta^*(L_t,n_{\text{ext}})$
on a  $64^3 \times 8$ lattice versus the strength of the external
chromomagnetic field, introduced on the lattice by constraining
the links according to eq.~(\ref{su2links}).
As in previous
section the critical coupling has been found by locating
the peak of the derivative of the free energy density with respect
to the gauge coupling $\beta$.
Figure~\ref{Fig5} shows our analysis for $T_c$ in units of $\Lambda_{\text{latt}}$
versus the critical temperature $a T_c$ together with a linear extrapolation
to the continuum. As one can ascertain there is evidence for a dependence of the critical
temperature on the applied field strength. As in the case of SU(3) the critical temperature can be expressed
in terms of a physical scale by using a parameterization for
the SU(2) string tension obtained by means of a fit
to the string tension data collected in Table~10 of ref.~\cite{Teper:1998kw}.
We interpolate the string tension data by
using Chebyshev polynomials of the first kind up to order 6 (see fig.~\ref{Fig5a}).

In fig.~\ref{Fig6}~  $T_c/\sqrt{\sigma}$ is plotted against
$\sqrt{gH}/\sqrt{\sigma}$. As in the SU(3) case discussed in previous section,
we can try to fit the data by means of a linear law.
Remarkably we found that the linear fit eq.~(\ref{tcsqrtsigma})
works quite well and we get
\be
\label{tcfitsu2}
\frac{T_c(0)}{\sqrt{\sigma}} = 0.710 (13) \qquad  \alpha = -0.126 (5)  \,.
\ee
The value obtained for $T_c(0)/\sqrt{\sigma}$ is in good agreement with the value
$T_c/\sqrt{\sigma}=0.694(18)$, without external field, obtained
in the literature~\cite{Teper:1998kw}.

Now we can estimate the critical field in string tension units that turns out to be
\be
\label{su2hc}
\frac{\sqrt{gH_c}}{\sqrt{\sigma}} = 5.33  \pm 0.33 \,.
\ee
Note that the critical field $\sqrt{gH_c}/\sqrt{\sigma}$ is about a factor 2 greater than the SU(3) critical value in eq.~(\ref{su3hc}).
This is at variance of the effective approach within dual superconductor picture in ref.~\cite{Chernodub:2002we},
where one gets  for the dual critical magnetic field $gH_c/\sigma=1$ for SU(2), while  $gH_c/\sigma=3/4$ for SU(3).
%
%
%
%
\FIGURE[ht]{\label{Fig7}
\includegraphics[width=0.85\textwidth,clip]{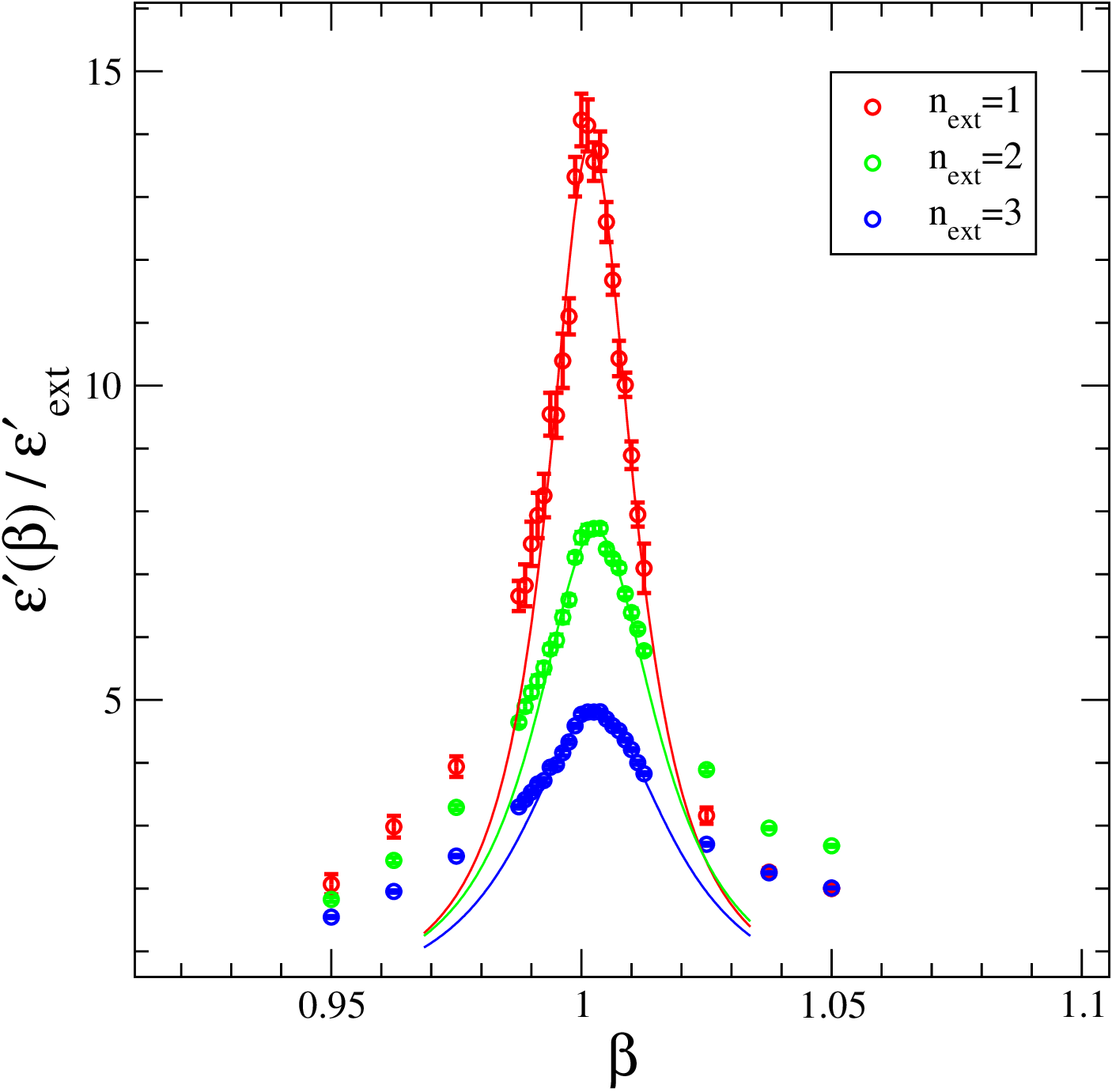}
\caption{U(1) in (3+1) dimensions. The derivative of the vacuum energy density with
respect to $\beta$, eq.~(\ref{avplaqu1}), versus $\beta$, for several values
of the strength of the constant magnetic background field on a $64 \times 16^3$ lattice. Solid lines
are the fits to the data near each of the peaks  using eq.~(\ref{peak-form}).
}
}
%
%
%
%
%
%
%
\FIGURE[ht]{\label{Fig8}
\includegraphics[width=0.85\textwidth,clip]{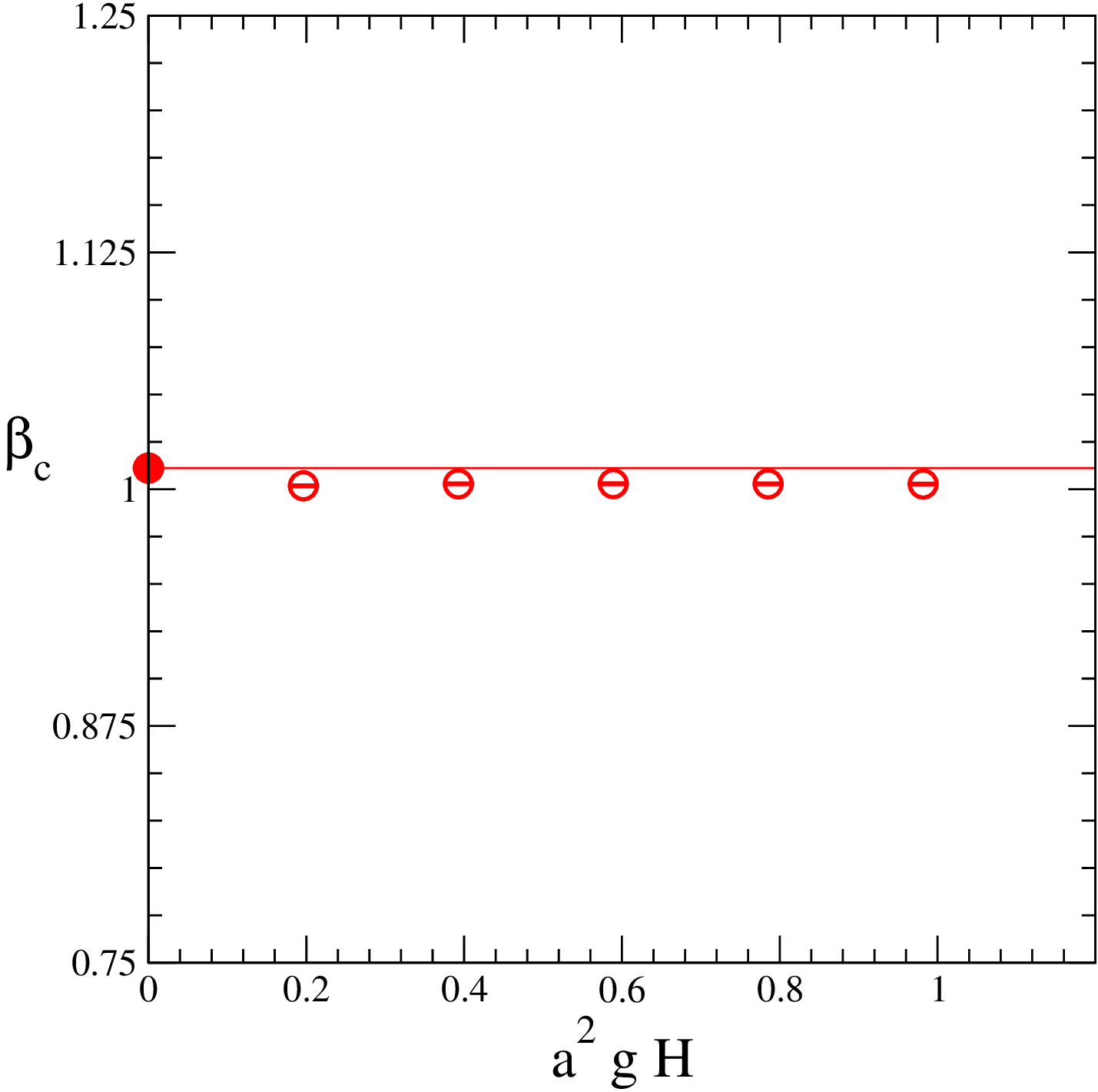}
\caption{U(1) in (3+1) dimensions. The critical coupling $\beta_c$ evaluated on a $64 \times 16^3$
lattice versus the strength of the constant background magnetic field (in lattice units).
The value at zero external field is the infinite volume extrapolation given in ref.~\cite{Arnold:2002jk}.
The solid line represents the central value of $\beta_c$ from ref.~\cite{Arnold:2002jk}.}
}
%
%

\subsection{U(1)}
\label{4dimU1}
In sections~\ref{4dimSU3} and~\ref{4dimSU2} we reported our results
indicating a dependence of the deconfinement temperature on the
strength of a constant abelian chromomagnetic background field. The
main aim of this section is to find out if the effect we found is
peculiar of non abelian gauge theories. To this purpose we consider
four dimensional U(1) lattice gauge theory.

It is known that, at zero temperature, U(1) lattice gauge theory
undergoes a weak first order phase
transition~\cite{Arnold:2002jk,Vettorazzo:2003fg,Vettorazzo:2004cr}
from the confined phase to the Coulomb phase for
$\beta=1.0111331(21)$ (using the standard Wilson action). We would
like to seek a possible dependence of the confinement-Coulomb phase
transition on the strength of an applied constant magnetic field.

The quantity we have measured to locate the critical coupling is
the derivative of the vacuum  energy density (with respect to the gauge coupling)
in presence of the background field (see sect.~\ref{gaugeinvariant})
\be
\label{avplaqu1}
\varepsilon^\prime(\beta,n_{\text{ext}}) = <U_{\mu\nu}>_{n_{\text{ext}}=0} -
<U_{\mu\nu}>_{n_{\text{ext}} \ne 0}  \,,
\ee
where $<U_{\mu\nu}>_{n_{\text{ext}}}$ is the average plaquette evaluated with
$n_{\text{ext}} \ne 0$ and $n_{\text{ext}}=0$ respectively.

In fig.~\ref{Fig7} we display the above quantity for three values of the
constant abelian background field, normalized to $\varepsilon^{\prime}_{\mathrm{ext}}$,
the derivative of the classical energy due to  the external applied field
\be
\label{epsprimeclass}
\varepsilon^{\prime}_{\mathrm{ext}} =
1 - \cos( a^2 g H ) = 1 -
\cos( \frac{2 \pi}{L_1} n_{\mathrm{ext}}) \,.
\ee
The values of $\beta$ corresponding to the peak in $\varepsilon^\prime(\beta,n_{\text{ext}})$
for several values of the strength of the applied constant abelian field are displayed in
fig.~\ref{Fig8}. Our conclusion is that, contrary to non abelian lattice gauge theories,
the critical coupling does not depend on the applied magnetic field strength.
Analogous result was found in ref.~\cite{Chernodub:2001da} for (2+1)-dimensional compact QED.

%
%
%
%
\FIGURE[ht]{\label{Fig9}
\includegraphics[width=0.85\textwidth,clip]{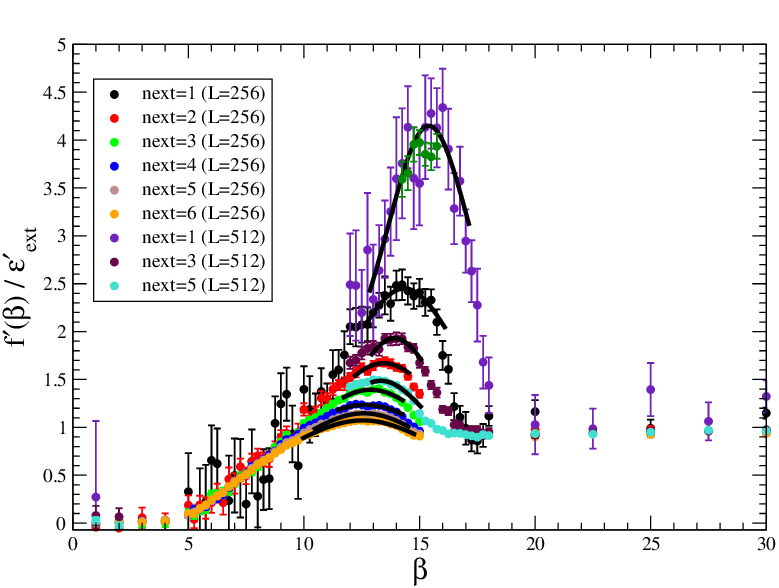}
\caption{SU(3) in (2+1) dimensions. The derivative of the free energy density with respect to
the gauge coupling $\beta$, eq.~(\ref{deriv}), versus $\beta$ for several values of the external field
strength. Lattice size is $L \times 256 \times L_t$ with two values of  $L=256,512$ and
temporal lattice size $L_t=4$. Solid lines are the fits eq.~(\ref{peak-form}).
}}
%
%
%
%
%
%
%
\FIGURE[ht]{\label{Fig10}
\includegraphics[width=0.85\textwidth,clip]{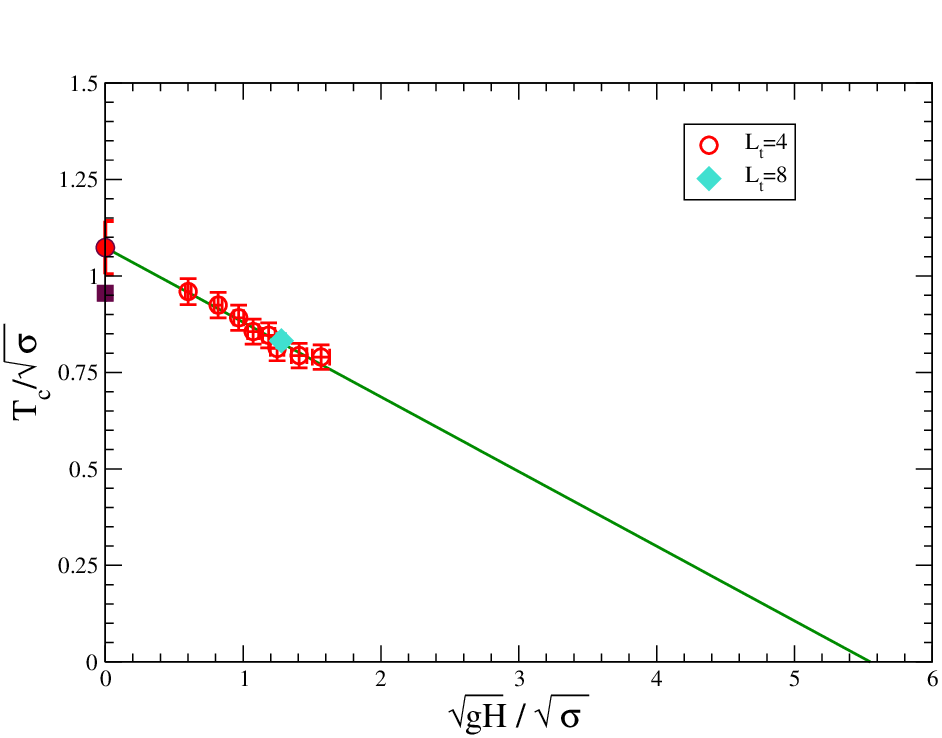}
\caption{SU(3) in (2+1) dimensions. The critical temperature $T_c$ estimated on  $256^2 \times 4$,
$512^2 \times 4$ and $512 \times 256  \times 8$ lattices  in units of
the string tension, eq.~(\ref{Tphys}), versus the square root of the field strength
$\sqrt{gH}$ in units of the string tension, eq.~(\ref{gHphys}). Open circles refer to $L_t=4$, diamond to $L_t=8$.
Solid line is the linear fit eq.~(\ref{tcsqrtsigma}).
In correspondence of $\sqrt{gH}/\sqrt{\sigma}=0$:  full circle  represents $T_c/\sqrt{\sigma}$ at zero external field obtained
by the linear extrapolation eq.~(\ref{tcsqrtsigma}), full square is the value given in ref.~\cite{Teper:1998te}.
}}
%
%
%
%
%
\FIGURE[ht]{\label{Fig11}
\includegraphics[width=0.85\textwidth,clip]{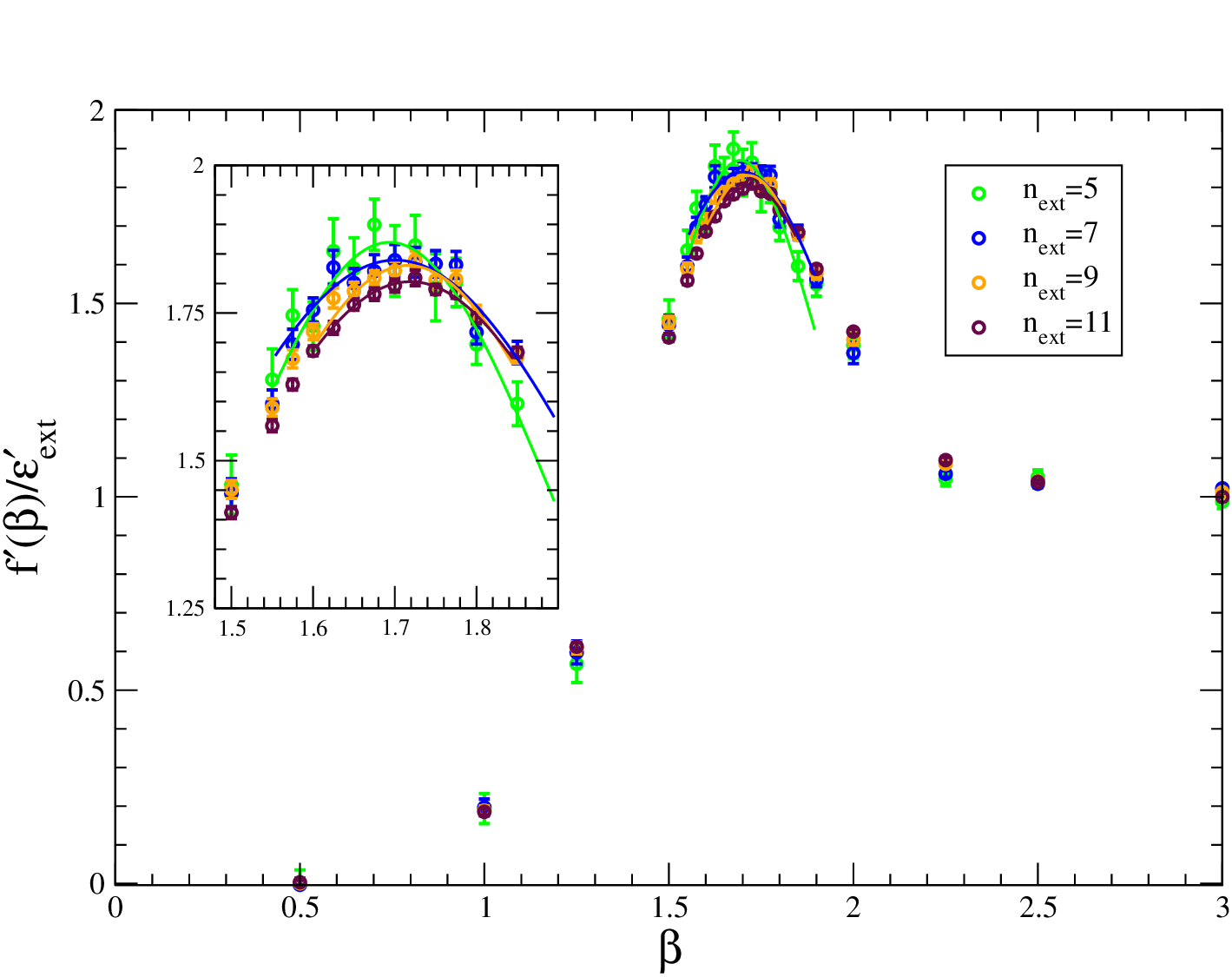}
\caption{U(1) in (2+1) dimensions. The derivative of the free energy density with
respect to $\beta$, eq.~(\ref{avplaqu1}), versus $\beta$ for several values
of the strength of the constant magnetic background field on a $512 \times 256 \times 4$ lattice. Solid lines
are the fits to the data near each of the peaks  using eq.~(\ref{peak-form}).
The inset is a magnification of the peak region.
}}
%
%
%
%
\FIGURE[ht]{\label{Fig12}
\includegraphics[width=0.85\textwidth,clip]{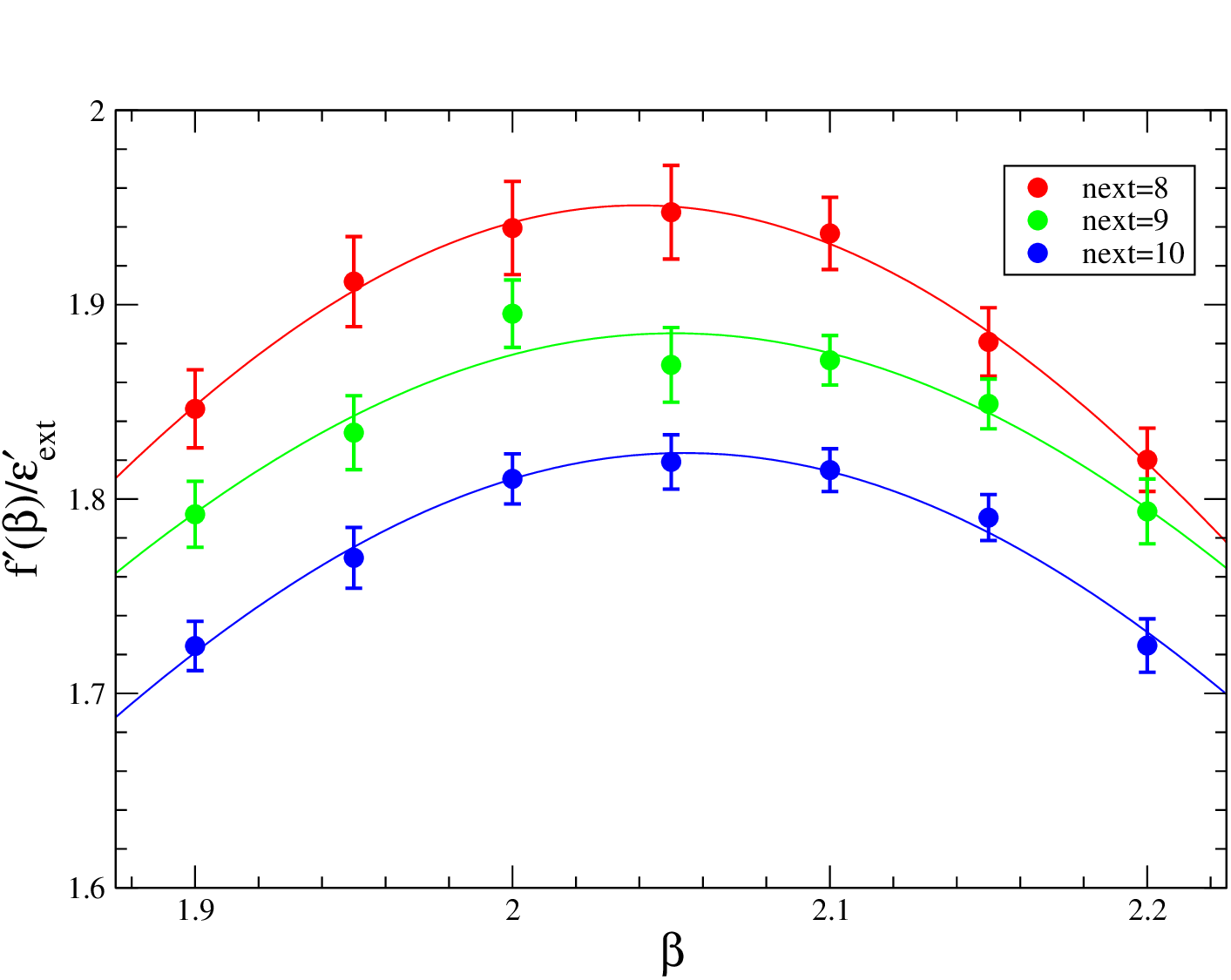}
\caption{U(1) in (2+1) dimensions. The derivative of the free energy density with
respect to $\beta$, eq.~(\ref{avplaqu1}), versus $\beta$ for several values
of the strength of the constant magnetic background field on a $512 \times 256 \times 8$ lattice. Solid lines
are the fits to the data near each of the peaks  using eq.~(\ref{peak-form}).
}}
%
%
%
%
\FIGURE[ht]{\label{Fig13}
\includegraphics[width=0.85\textwidth,clip]{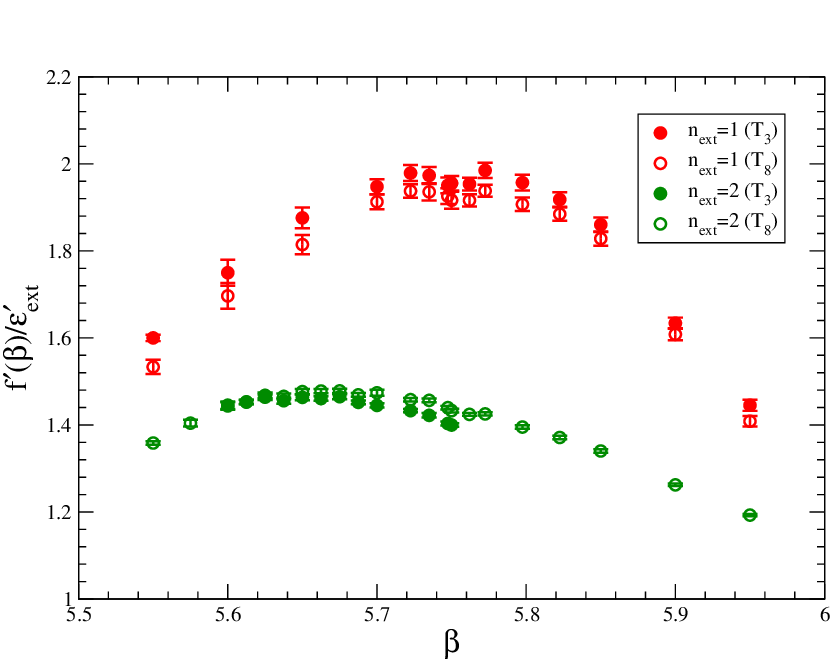}
\caption{SU(3) in (3+1) dimensions. The derivative of the free energy density with respect to
the gauge coupling $\beta$, eq.~(\ref{deriv}), versus $\beta$ at fixed external field
strength ($n_{\text{ext}}=1$) for a $64^3 \times 8$ lattice.
Open symbols refer to $T_3$ abelian chromomagnetic field, induced by lattice links in eq.~(\ref{t3links}).
Full symbols refer to $T_8$ abelian chromomagnetic field eq.~(\ref{t8links}).
}}
%

\section{(2+1) dimensions}
\label{3dim}
Our numerical results for non abelian gauge theories SU(2) and SU(3) in (3+1) dimensions
in presence of an abelian constant chromomagnetic background field lead us to conclude that the
deconfinement temperature depends on the strength of the applied field, and eventually
becomes zero for a critical value of the field strength.
A natural question arises if this phenomenon, which is peculiar of non abelian gauge theories, continues
to hold in (2+1) dimensions. To this purpose we consider here the non abelian SU(3) lattice gauge theory
to be contrasted with the abelian U(1) lattice gauge theory at finite temperature.

\subsection{SU(3)}
\label{3dimSU3}
In this section we focus on gauge systems in (2+1) dimensions.
As is well known gauge theories in (2+1) dimensions possess a dimensionful coupling constant, namely
$g^2$ has dimension of mass and so provides a physical scale.

Non abelian gauge theories in (2+1)  and (3+1) dimensions are sufficiently similar.
Indeed, lattice simulations provide convincing evidence that (2+1) dimensional
SU(N) gauge theories confine with a linear potential~\cite{Teper:1998te}.
Moreover, at finite temperature there is a deconfinement
transition~\cite{Engels:1997dz}.\\
In (2+1) dimensions the chromomagnetic field $H^a$
is a (pseudo)scalar
\be
\label{B3dim}
H^a = \frac{1}{2} \varepsilon_{ij} F^a_{ij} = F^a_{12}   \,.
\ee
For SU(3) gauge theory a constant abelian chromomagnetic field $H^3$
can be obtained with
\be
\label{3dlinks}
\begin{split}
& U^{\mathrm{ext}}_1(\vec{x}) = {\mathbf{1}} \,,
\\
& U^{\mathrm{ext}}_2(\vec{x}) =
\begin{bmatrix}
\exp(i \frac {g H x_1} {2})  & 0 & 0 \\ 0 &  \exp(- i \frac {g H
x_1} {2}) & 0
\\ 0 & 0 & 1
\end{bmatrix}
\end{split}
\ee
As in the four dimensional case (see sect.~\ref{abelianfield})
since we assume to have a lattice with toroidal geometry the field strength is quantized
\be
\label{3dimquant}
a^2 \frac{g H}{2} =
\frac{2 \pi}{L_1} n_{\mathrm{ext}} \,, \qquad  n_{\mathrm{ext}}\,\,\,{\text{integer}} \,.
\ee
We computed the derivative of the free energy density eq.~({\ref{freeenergy}) on a $L \times 256 \times 4$ lattice,
with $L=256, 512$ and several values of the external field strength parameterized by $n_{\text{ext}}$.
Our numerical results are reported in fig.~\ref{Fig9}.
We locate the critical coupling $\beta_c$ as the position of the maximum of the derivative of the free energy density
at given external field strength. As for SU(3) in (3+1) dimensions, the value of $\beta_c$ depends on the
field strength. Using the parameterization for the string tension given in eq.~(C9) of ref.~\cite{Teper:1998te}
\be
\label{su3string3dim}
\beta a \sqrt{\sigma} = 3.367(50) + \frac{4.1(1.7)}{\beta} + \frac{46.5(11.0)}{\beta^2}
\ee
we are able to estimate the critical temperature $T_c$ in units of the string tension.
We find that, as in (3+1) dimensions,  $T_c/\sqrt{\sigma}$ depends linearly
on the applied field strength (see fig.~\ref{Fig10}).
The linear fit eq.~(\ref{tcsqrtsigma}) gives
\be
\label{su3Tc3dim}
\frac{T_c(0)}{\sqrt{\sigma}}= 1.073 (87) \quad \alpha=-0.193(76) \,,
\ee
that implies a critical field in string tension units
$\sqrt{gH_c}/\sqrt{\sigma} = 5.5 \pm 3.7$.
Note that value for $T_c(0)/\sqrt{\sigma}$ in the present work is in fair agreement
with $T_c/\sqrt{\sigma}=0.972(10)$ without external field obtained in ref.~\cite{Engels:1997dz}.
To check possible finite volume effects, we performed a lattice simulation with $L_t=8$.
The result, displayed in fig.~\ref{Fig10}, shows that within statistical uncertainties
our estimate of the critical temperature from the simulation with $L_t=8$ is in agreement
with result at $L_t=4$.
%
%

\subsection{U(1)}
\label{3dimU1}
In a classical paper~\cite{Polyakov:1976fu} Polyakov showed that compact quantum electrodynamics in (2+1) dimensions
at zero temperature confines external charges for all values of the coupling. Moreover
it is well ascertained that the confining mechanism is the condensation of
magnetic monopoles which gives rise to a linear confining potential and a non-zero string tension
\be
\label{polstring}
\sigma a^2 = \kappa \frac{1}{\sqrt{\beta}} \exp [-\pi^2 V(0) \beta]
\ee
where $\kappa$ is a constant, $V(0)=0.2527$~\cite{Banks:1977cc} is the value of the lattice propagator
at zero separation, and $\beta = 1/(a g^2)$.

At finite temperature it is well known that the gauge system undergoes a deconfinement transition which appears
to be of the Kosterlitz-Thouless type~\cite{Coddington:1986jk}.
We are interested in lattice U(1) gauge theory in an uniform external magnetic field
\be
\label{u1links3d}
\begin{split}
& U^{\text{ext}}_1(\vec{x}) =  {\mathbf{1}} \,, \\
& U^{\text{ext}}_2(\vec{x}) = \cos(  gHx_1) +
i \sin( gHx_1 )   \,.
\end{split}
\ee
We performed numerical simulations on $512 \times 256 \times 4$ and $512 \times 64 \times 8$
lattices. We measure the derivative of the free energy density with respect to the coupling $\beta$.
In fig.~\ref{Fig11} we display the results for the $512 \times 256 \times 4$ lattice.
To determine the critical coupling $\beta_c$, we fitted the lattice data to eq.~(\ref{peak-form}).
Contrary to the case of (2+1) and (3+1) non abelian lattice gauge theories, we do not find
a dependence of the critical value of the coupling $\beta_c$ on the magnetic field strength.
Indeed we found that (temporal size $L_t=4$)
\be
\label{betacu13dlt4}
\begin{split}
 \beta_c(n_{\text{ext}}=5)  &=  1.694(17)  \\
 \beta_c(n_{\text{ext}}=7)  &=  1.701(13)  \\
 \beta_c(n_{\text{ext}}=9)  &=  1.716(11)  \\
 \beta_c(n_{\text{ext}}=11) &=  1.719(10)) \\
\end{split}
\ee
By increasing the temporal size to $L_t=8$ the critical coupling increases and is still independent
of the external magnetic field strength (see fig.~\ref{Fig12}). Indeed,  we found
\be
\label{betacu13dlt8}
\begin{split}
 \beta_c(n_{\text{ext}}=8)   &=  2.040(14)  \\
 \beta_c(n_{\text{ext}}=9)   &=  2.051(15)  \\
 \beta_c(n_{\text{ext}}=10)  &=  2.054(12)  \\
\end{split}
\ee
Therefore we can conclude that even in (2+1) dimensional case the critical coupling does not depend
on the strength of the external magnetic field as for U(1) lattice gauge theories in (3+1) dimensions (see sect.~\ref{4dimU1}).

%
\section{Conclusions}
\label{Conclusions}
Let us conclude this paper by stressing our main results.
We have investigated U(1), SU(2), and SU(3) pure gauge theories both in
(3+1) and (2+1) dimensions in presence of an uniform (chromo)magnetic field.
For non abelian gauge theories we found that there is a critical field
$gH_c$ such that for $gH > gH_c$ the gauge systems are in the deconfined phase.
Moreover, such an effect seems to be generic for non abelian gauge theories.
On the other hand our numerical results for abelian gauge theories, where it
is well established~\cite{Polyakov:1976fu,Fradkin:1978th} that confinement is due
to monopole condensation, do not show any dependence of the critical coupling
from the strength of an external magnetic field. Therefore it seems very difficult
to explain our reversible color Meissner effect in SU(2) and SU(3) gauge theories
in terms of abelian color magnetic monopoles. Instead, the peculiar dependence of the
deconfinement temperature on the strength of the abelian chromomagnetic field
$gH$ could be naturally explained if the vacuum behaved as an ordinary relativistic
color superconductor, namely a condensate of color charged scalar fields whose
mass is proportional to the inverse of the magnetic length. However, the chromomagnetic
condensate cannot be uniform due to gauge invariance of the vacuum, which disorders the
gauge system in such a way that there are not long range correlations.
Consequently  we can speculate that if the vacuum behaved as a non uniform chromomagnetic condensate,
our reversible color Meissner effect
could be easily explained, for strong enough chromomagnetic fields would force long
range color correlations such that the gauge system gets deconfined.
One might thus imagine the confining vacuum in non abelian gauge systems as a disordered
chromomagnetic condensate which confines color charges due both to the presence of
a mass gap and the absence of long range color correlations, as
argued by R.P. Feynman in (2+1) dimensions~\cite{Feynman:1981ss}.

\appendix

%
\section{SU(3) $T_8$ abelian chromomagnetic background field}
\label{su3T8}

As is well known, in SU(3) there are two independent ways to realize
a constant abelian chromomagnetic field.
The first one, that we have considered in section~\ref{4dimSU3}, is to take the
abelian field directed along direction $\hat{3}$ in color space; the
second one is to take the field along direction  $\hat{8}$ in color
space.
In this Appendix we consider SU(3) lattice gauge theory in
(3+1) dimensions in presence of a constant chromomagnetic background
field along direction $\hat{8}$ in SU(3) color space and along
spatial direction $\hat{3}$. The continuum gauge potential and the
corresponding lattice links are defined in eq.~(\ref{su3pot}) and
eq.~(\ref{t8links}) respectively.
One should not expect a vastly different behavior in the two cases.
Indeed we found that, even for a constant abelian chromomagnetic
background field directed along color direction $\hat{8}$, the
critical deconfinement temperature $T_c$ depends on the strength of
the applied field.
Even more fig.~\ref{Fig13}, where we  display the derivative of the free energy density with
respect to the gauge coupling $\beta$ for $n_{\text{ext}}=1,2$ on a
$64^3 \times 8$ lattice, shows that  the derivative of the free energy
density in presence of the abelian chromomagnetic background field directed
along color space direction $\hat{8}$
behaves like the case in which the field is directed along color space direction $\hat{3}$.
Moreover the critical couplings at fixed field strength are consistent within statistical errors.


\providecommand{\href}[2]{#2}\begingroup\raggedright\endgroup

\end{document}